\documentclass[reprint,aps,nofootinbib,twocolumn,amsmath,amssymb]{revtex4-1}
\usepackage{graphicx}
\usepackage{stmaryrd}
\usepackage{amsfonts}
\usepackage{mathrsfs}
\usepackage{color}
\usepackage{bm}
\usepackage{amsmath}
\usepackage{chemarrow}
\def\dbar{{\mathchar '26\mkern -10mu\delta}}
\def\overline{\underline}

\begin{document}
\title{Covariant phase space with null boundaries}

\author{Kai Shi$^1$}
\thanks{kaishi@mail.bnu.edu.cn}
\author{Xuan Wang$^1$}
\thanks{xuanwang@mail.bnu.edu.cn}
\author{Yihong Xiu$^1$}
\thanks{yihongxiu@mail.bnu.edu.cn}
\author{Hongbao Zhang$^{1,2}$}
\thanks{hzhang@vub.ac.be}

\affiliation{$^1$Department of Physics, Beijing Normal University, Beijing 100875, China\\
$^2$Theoretische Natuurkunde, Vrije Universiteit Brussel, and The International Solvay
Institutes, Pleinlaan 2, B-1050 Brussels, Belgium}

\begin{abstract}
By imposing the boundary condition associated with the boundary structure of the null boundaries rather than the usual one, we find for Einstein's gravity that the variational principle works only in its submanifold with the null boundaries given by the expansion free and shear free hypersurfaces rather than in the whole covariant phase space. This implies that the key requirement in Harlow-Wu's algorithm for the timelike boundaries is too restrictive for the null ones. To incorporate more generic situations into Harlow-Wu's algorithm, we relax such a requirement. As a result, we successfully reproduce the Hamiltonian obtained previously by Wald-Zoupas' prescription for Einstein's gravity.
\end{abstract}



\maketitle

\section{Introduction and motivation}
Developed mainly by Wald and his companions in \cite{LW,Wald1,IW}, not only does the covariant phase space formalism give a covariant way to understand the Hamiltonian dynamics of classical field theories, but also provides an elegant avenue towards the quantum counterpart of classical field theories\footnote{The readers are also referred to \cite{BB,BC,BT} for the alternative formalism in terms of surface charge algebra.}. Moreover, the covariant phase space formalism plays an important role in investigating the contribution to the entanglement entropy from the edge modes\cite{DF,Speranza}. In particular, based mainly on the consistency condition in mathematics that the exterior derivative of an exact form in configuration space should vanish and the reasonable assumption in physics that the flux on stationary solutions should vanish, Wald and Zoupas present a proposal for the definition of the Hamiltonians associated with infinitesimal asymptotic symmetries within the covariant phase space framework, which agree with those obtained previously from entirely different considerations\cite{WZ}. But nevertheless, Wald-Zoupas' prescription treats the boundary terms in a cavalier manner, leading to the potential difficulty in applying it to a variety of circumstances. Gratefully, such an annoying situation has recently been ameliorated by Harlow and Wu in \cite{HW}, where guided by the variational principle, they propose an algorithm to reproduce the aforementioned Hamiltonians in a direct way by taking into account the boundary terms in a meticulous manner. Such an algorithm  is further formalized into the relative bicomplex framework in \cite{MV} later on.

 However, the boundaries considered in \cite{HW} are timelike. Accordingly, the resulting Hamiltonian gives rise to the ADM mass at the spatial infinity of an asymptotically flat spacetime by pushing the location of the timelike boundary to infinity. This corresponds essentially to Case I of \cite{WZ}. While Case II of \cite{WZ} involves null boundaries, such as the null infinity in asymptotically flat spacetimes considered in \cite{WZ} and the null hypersurface at a finite location considered in \cite{CFP}. Therefore not only is it tempting but also important to check whether Harlow-Wu's algorithm gives rise to the same result as obtained by Wald-Zoupas' prescription associated with null boundaries. As a first step along this line, in this paper we focus exclusively on the null boundaries at a finite location, which have recently received a lot of attentions\cite{DGGP1,HPS1,DGGP2,HPS2,HF1,HHPS,HF2,DGGP,DPS,DM,DG,HPS3,RW,AGSSZ,DGO,GPSTZ,GSZ,ASTYZ}. As a result, we find that the key requirement in Harlow-Wu's algorithm for the pre-symplectic potential at the timelike boundaries as an exact form in configuration space plus an exact form in spacetime manifold, or equivalently to require the variational principle be well defined, is too restrictive, in particular for the null ones. By relaxing such a requirement as to allow the presence of an additional flux term in the canonical form, we successfully reproduce the Hamiltonian obtained previously by Wald-Zoupas' prescription for Einstein's gravity with the null boundaries at a finite location, although it is not conserved in general. We believe that our relaxation of Harlow-Wu's algorithm has been as general as possible such that it can incorporate all the circumstances one may encounter.

The structure of this paper is organized as follows. In the next section, we shall review the basic notions of null hypersurfaces, where we define free horizons as expansion free and shear free null hypersurfaces. In Section \ref{ba}, we further introduce the boundary structure of null hypersurfaces and its associated asymptotic symmetries. Then in the subsequent section, by imposing the boundary condition associated with the aforementioned boundary structure at the null boundaries, we show that the variational principle for Einstein's gravity does not work in the whole covariant phase space. Instead, it works in its submanifold consisting of the free horizons as the null boundaries. In Section \ref{competition}, after relaxing the key requirement in Harlow-Wu's algorithm such as to incorporate more generic situations, we show that the Hamiltonians obtained previously by Wald-Zoupas' prescription can be successfully reproduced from scratch by Harlow-Wu's algorithm. We conclude our paper with some discussions in the last section.

For the most part, we shall follow the conventions and notations in \cite{Wald2} with the mostly pluses signature except that we set $16\pi G=1$.
\section{Null hypersurfaces and free horizons}
\begin{figure}
\begin{center}
  \includegraphics[scale=0.42]{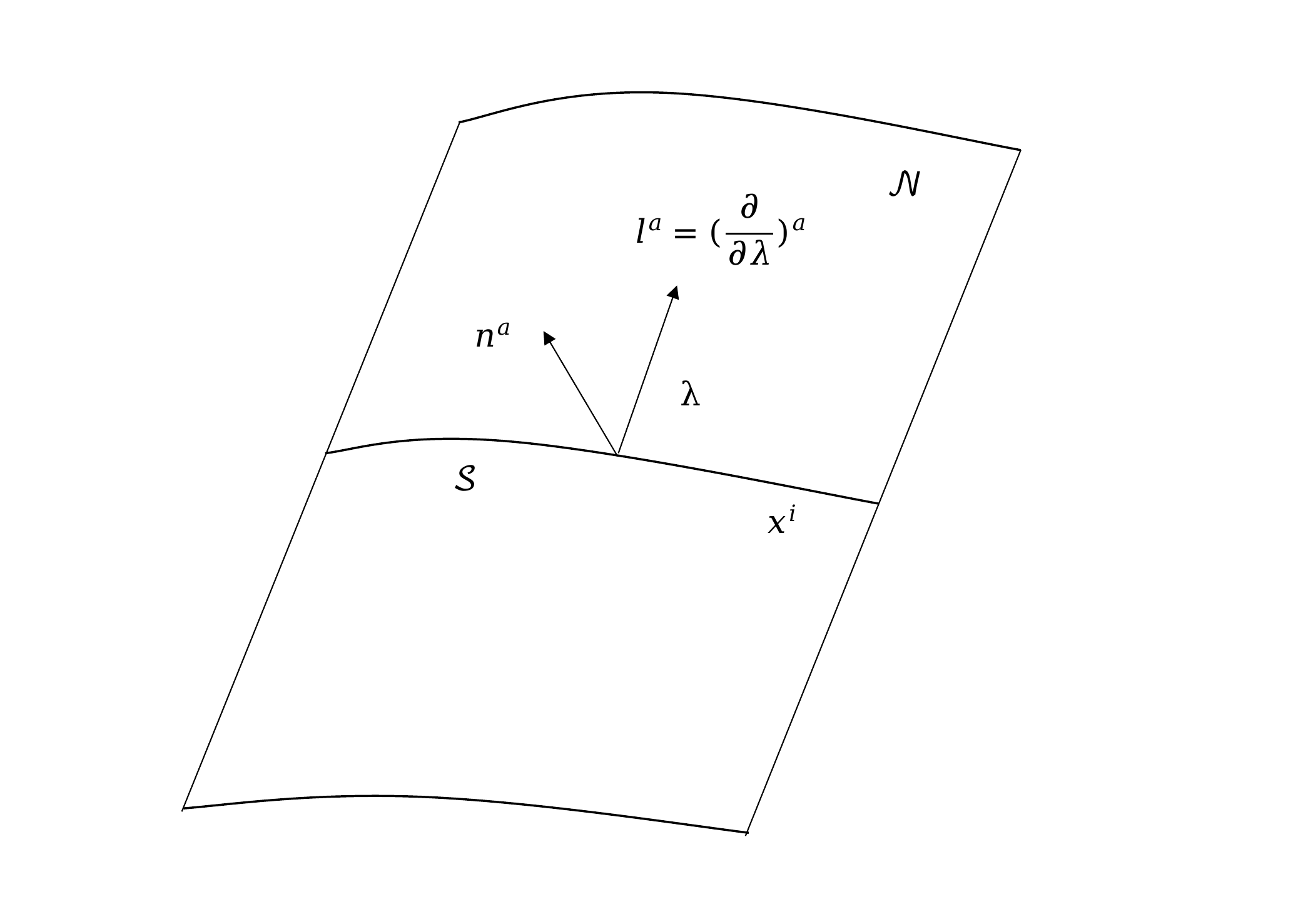}
  \end{center}
  \caption{The foliation of the null hypersurface $\mathcal{N}$ by the cross-section $\mathcal{S}$ and the null generators $l^a$ gives rise to a natural coordinate system $\{\lambda,x^i\}$ on $\mathcal{N}$, with the induced volume $\overline{\bm{\epsilon}}^{(d-1)}=\sqrt{q}\text{d}\lambda \text{d}^{d-2}x$ on $\mathcal{N}$ and  $\overline{\bm{\epsilon}}^{(d-2)}=\sqrt{q} \text{d}^{d-2}x$ on $\mathcal{S}$.}\label{hypersurface}
\end{figure}

A codimension-one submanifold $\mathcal{N}$ in a $d$-dimensional spacetime $(\mathcal{M},g_{ab})$ is called a null hypersurface if its normal $l_a$ satisfies
\begin{equation}\label{nhd}
l_al^a\;\widehat{=}\; 0, \quad \text{d}\bm{l}\;\widehat{=}\;\bm{w}\wedge \bm{l},
\end{equation}
where $\widehat{=}$ means the equation evaluated at $\mathcal{N}$ and the bolded letters denote differential forms with $\bm{w}$ depending on how $\bm{l}$ extends off $\mathcal{N}$. Below we require $l^a$ be future-directed and assume that $\mathcal{N}$ is diffeomorphic to $\mathcal{Z}\times\mathbb{R}$ with $\mathcal{Z}$ the manifold of integral curves of null generators.
At each point $p\in \mathcal{N}$, one can have a natural subspace $V_p(\mathcal{N})$ of $V_p(\mathcal{M})$, whose element is tangent to $\mathcal{N}$  as $v^al_a=0$. Furthermore, note that $l^a$ is tangent to $\mathcal{N}$, so one can define an equivalence class space of vectors $\hat{V}_p(\mathcal{N})$,  where the two vectors tangent to $\mathcal{N}$ are equivalent if they differ by a multiple of $l^a$.  On the other hand, the image of the pullback of $V_p^*(\mathcal{M})$ to $\mathcal{N}$ comprises $V_p^*(\mathcal{N})$, which is also isomorphic to the equivalence class space of dual vectors where two dual vectors are equivalent if they differ by a multiple of $l_a$ due to the fact that
 \begin{equation}
 \overline{l}_a=0
 \end{equation}
 where the underline denotes the pullback to $\mathcal{N}$.  Furthermore, the image of the pullback of the subspace of $V^*_p(\mathcal{M})$ consisting of those vectors satisfying $\overline{u}_al^a=u_al^a=0$ has a natural correspondence with $\hat{V}^*_p(\mathcal{N})$.
More generally, a tensor $T$ over $V_p(\mathcal{M})$ can naturally give rise to a tensor $\hat{T}$ over $\hat{V}_p(\mathcal{N})$ as
 \begin{equation}
\hat{ T}^a{}_b \hat{u}_a\hat{v}^b=T^a{}_bu_av^b
 \end{equation}
 with $u_al^a=v^al_a=0$ if and only if the result of contracting any one of its indices with $l_a$ or $l^a$ and contracting its remaining indices with vectors $v^a$ or dual vectors $w_a$ vanishes. Obviously, the tensor product of $l^a$ or $l_a$ with any other tensor gives rise to a vanishing hatted tensor.
By the aid of the basis as
 \begin{equation}
 (e_+)^a=l^a,\quad (e_-)^a=n^a, \quad (e_i)^a, \quad i=1,\cdots,d-2
 \end{equation}
 with $n^al_a=-1$ and $l_a(e_i)^a=n_a(e_i)^a=0$, one can readily show that a tensor can be hatted if and only if it can be expressed as a tensor over the subspace spanned by $\{(e_i)^a\}$ plus a summation of tensor products of $l^a$ or $l_a$ with other tensors.
  In addition, there is a natural one-to-one correspondence between the hatted tensors and the tensors over the subspace spanned by $\{(e_i)^a\}$. So two tensors give rise to the same hatted tensor if and only if they differ by a summation of tensor products of $l^a$ or $l_a$ with other tensors. On the other hand, note that the pullback of a tensor which can be hatted also has a natural correspondence with a tensor over the subspace spanned by $\{(e_i)^a\}$, so such a pullback can be identified with a hatted tensor. It is noteworthy that with a choice of the cross-section $\mathcal{S}$, the vector space tangent to $\mathcal{S}$ naturally gives rise to such a subspace. Moreover, together with the null generators parameterized as $l^a=(\frac{\partial}{\partial \lambda})^a$, such a cross-section also gives a foliation of $\mathcal{N}$, which is depicted in Figure.\ref{hypersurface}.

Now let us introduce some geometric quantities induced on $\mathcal{N}$ from the spacetime. The first one is the induced metric $\overline{g}_{ab}$, which is obviously degenerate because
\begin{equation}
\overline{g}_{ab}l^b=0,
\end{equation}
where the underlined operation denotes the restriction onto $\mathcal{N}$.
The second one is the induced volume, which is defined as
\begin{equation}\label{iv}
-\bm{l}\wedge\bm{\epsilon}^{(d-1)}=\bm{\epsilon}^{(d)}
\end{equation}
with $\bm{\epsilon}^{(d)}$ the volume element associated with $g_{ab}$ in $\mathcal{M}$.
Whence the divergence of a vector field tangent to $\mathcal{N}$ can be uniquely defined as
\begin{equation}
D_av^a\overline{\bm{\epsilon}}^{(d-1)}=\mathscr{L}_v\overline{\bm{\epsilon}}^{(d-1)}=\text{d}(v\cdot\overline{\bm{\epsilon}}^{(d-1)}),
\end{equation}
where the covariant derivative operator $D_a$ is only required to be compatible with $\overline{\bm{\epsilon}}^{(d-1)}$ and the dot denotes the contraction of $v^a$ with the first index of $\overline{\bm{\epsilon}}^{(d-1)}$. In passing, one can also define the induced volume on a cross-section $\mathcal{S}$ as
\begin{equation}
\bm{\epsilon}^{(d-2)}=l\cdot\bm{\epsilon}^{(d-1)}.
\end{equation}
The third one is the so-called surface gravity $\kappa$, characterizing the non-affinity of the null geodesics generated by $l^a$, which is defined through the following equation
\begin{equation}\label{sg}
l^a\nabla_al^b=\frac{1}{2}\nabla^b(l^al_a)+l^aw_al^b=\kappa l^b,
\end{equation}
where Eq. (\ref{nhd}) has been used.  The fourth one is the second fundamental form $K_{ab}=\overline{\nabla_al_b}$, which satisfies
\begin{equation}
K_{ab}=K_{(ab)},\quad K_{ab}=\frac{1}{2}\mathscr{L}_l\overline{g}_{ab}, \quad l^aK_{ab}=0,
\end{equation}
where we have used Eq. (\ref{nhd}) for the first equation and the fact that the Lie derivative commutes with the pullback for the second equation.
The fifth one is the so-called  Weingarten map $W_a{}^b=\overline{\nabla}_al^b$, which is well defined because
\begin{equation}
W_a{}^bl_b=0.
\end{equation}
It is noteworthy that both the surface gravity and second fundamental form can be obtained from the Weingarten map as
\begin{equation}
l^aW_a{}^b=\kappa l^b,\quad K_{ab}=W_a{}^c\overline{g}_{cb}.
\end{equation}

A free horizon is a null hypersurface with a vanishing second fundamental form as $K_{ab}=0$, implying the invariance of the metric along the null generators. With a free horizon, one has
\begin{equation}
W_a{}^b=\omega_al^b, \quad \kappa=l^a\omega_a
\end{equation}
where $\omega_a$ is called the rotation one-form. Note that one can decompose
\begin{equation}
\widehat{\nabla_al_b}=\frac{1}{d-2}\theta\hat{g}_{ab}+\hat{\sigma}_{ab}
\end{equation}
with $\theta$ the expansion and $\hat{\sigma}_{ab}$ the traceless shear of null generators, respectively. So a null hypersurface is a free horizon if and only if it is expansion free and shear free, which explains our definition. By expanding the metric in terms of the previous basis as
\begin{equation}
g_{ab}=-l_an_b-l_bn_a+q_{ab}
\end{equation}
with $q_{ab}$ the tensor over the subspace spanned by $\{(e_i)^a\}$, one can show that
\begin{eqnarray}
\theta&=&\hat{g}^{ab}\widehat{\nabla_al_b}=\hat{q}^{ab}\widehat{\nabla_al_b}=q^{ab}\nabla_al_b\nonumber\\
&=&\frac{1}{2}q^{ab}\mathscr{L}_lg_{ab}=\frac{1}{2}q^{ab}\mathscr{L}_lq_{ab},
\end{eqnarray}
where we have used
\begin{equation}\label{Lie}
\mathscr{L}_v\bm{l}=\text{d}(v\cdot \bm{l})+v\cdot \text{d}\bm{l}=w_v\bm{l}
\end{equation}
for a vector field $v^a$ tangent to $\mathcal{N}$ in the last step. On the other hand, one can also express the expansion as
\begin{eqnarray}
\theta&=&q^{ab}\nabla_al_b=(g^{ab}+l^an^b+l^bn^a)\nabla_al_b\nonumber\\
&=&\nabla_al^a+n\cdot\mathscr{L}_l\bm{l}=\nabla_al^a-w_l=D_al^a,
\end{eqnarray}
where we have used Eq. (\ref{iv}) in the last step.  By Eq. (\ref{sg}) and Eq. (\ref{Lie}), the covariant derivative and Lie derivative of a hatted tensor along $l$ can be well defined as
\begin{equation}
l^c\nabla_c\hat{T}^a{}_b=\widehat{l^c\nabla_c T^a{}_b},\quad \mathscr{L}_l\hat{T}^a{}_b=\widehat{\mathscr{L}_lT^a{}_b},
\end{equation}
which is linear, satisfying the Leibniz law, and commuting with the contraction between the two hatable tensors.
In particular, by hatting
\begin{equation}
l^c\nabla_c\nabla_al_b=\nabla_a\kappa l_b+\kappa\nabla_al_b-\nabla_al^c\nabla_cl_b-R_{cadb}l^cl^d,
\end{equation}
we arrive at the Raychaudhuri equation as follows
\begin{eqnarray}
l^c\nabla_c\theta&=&\kappa\theta-\frac{1}{d-2}\theta^2-\hat{\sigma}_{ab}\hat{\sigma}^{ab}-R_{cd}l^cl^d,\nonumber\\
l^c\nabla_c\hat{\sigma}_{ab}&=&-\theta\hat{\sigma}_{ab}-\widehat{C_{cadb}l^cl^d}.
\end{eqnarray}
If $R_{cd}l^cl^d\geq0$, which can be achieved by the reasonable null energy condition in Einstein's general relativity, then the expansion free implies the shear free. In this case, a free horizon is a non-expanding horizon in \cite{ABF1,ACK,ABF2,AFK,ABDFK,ABL,AK}.

\section{Boundary structure of null hypersurfaces and asymptotic symmetries at null boundaries\label{ba}}
In this section, we shall provide a brief review of the boundary structure of null hypersurfaces and the corresponding asymptotic symmetries. For more details, please refer to \cite{CFP}.

Given a null hypersurface $\mathcal{N}$ as a boundary of our theory in $(\mathcal{M},g_{ab})$, one can define the induced boundary structure $\mathfrak{p}$ of $\mathcal{N}$ as the equivalence class of a triple $\left(l^a,\kappa,l_a\right)$ with the equivalence relation given by
\begin{equation}\label{triple}
\left(l^a,\kappa,l_a\right)\sim\left(l'^a=e^\alpha l^a,\kappa'=e^\alpha(\kappa+\mathscr{L}_l\alpha), l'_a=e^\alpha l_a\right),
\end{equation}
where $\alpha$ is a smooth function on $\mathcal{N}$. One can further obtain the universal structure $\mathfrak{u}$ of $\mathcal{N}$ when restricted to the equivalence class of the pair $\left(l^a,\kappa\right)$.

The vector field $\chi$ tangent to $\mathcal{N}$ generates the diffeomorphisms on $\mathcal{N}$ that preserve the above universal intrinsic structure $\mathfrak{u}=\left[l^a,\kappa\right]$ if it obeys
\begin{equation}
\mathscr{L}_\chi l^a=\beta l^a,\quad \mathscr{L}_\chi\kappa=\kappa\beta+\mathscr{L}_l\beta
\end{equation}
with $\beta$ satisfying the following normalization condition
\begin{equation}
\beta(\chi^a,e^\alpha l^a)=\beta(\chi^a,l^a)+\mathscr{L}_\chi\alpha.
\end{equation}
These vector fields form a Lie algebra $\mathfrak{g}_\mathfrak{u}$ of infinitesimal symmetry group, where  the Lie bracket is given by the commutator with
\begin{equation}
\beta([\chi_1,\chi_2]^a,l^a)=\mathscr{L}_{\chi_1}\beta(\chi_2^a,l^a)-\mathscr{L}_{\chi_2}\beta(\chi_1^a,l^a).
\end{equation}
As detailed in \cite{CFP}, the symmetry algebra has the structure
\begin{equation}
\mathfrak{g}_\mathfrak{u}\simeq\text{diff}(\mathcal{Z})\ltimes\mathfrak{s}\simeq\text{diff}(\mathcal{Z})\ltimes(\mathfrak{b}\ltimes\mathfrak{s}_0),
\end{equation}
where $\ltimes$ denotes the semidirect sum, $\mathfrak{s}$ is the algebra of generalized supertranslations $fl^a$ with
\begin{equation}
\mathscr{L}_l(\mathscr{L}_l+\kappa)f=0,
\end{equation}
$\mathfrak{s}_0$ is the algebra of affine supertranslations $fl^a$ with
\begin{equation}
(\mathscr{L}_l+\kappa)f=0,
\end{equation}
and $\mathfrak{b}$ is the quotient algebra $\mathfrak{s}/\mathfrak{s}_0$.
Similarly, one can construct the symmetry algebra consisting of the vector fields on $\mathcal{M}$ which preserves the boundary $\mathcal{N}$ and the boundary structure $\mathfrak{p}$ as follows
\begin{equation}
\mathfrak{h}_\mathfrak{p}=\{\xi^a|\xi^a\;\widehat{=}\;\chi^a, \ \gamma\;\widehat{=}\;\beta\}
\end{equation}
with $\gamma$ given by
\begin{equation}
\mathscr{L}_\xi l_a\;\widehat{=}\;\gamma l_a.
\end{equation}
To obtain the asymptotic symmetry algebra, one is required to factor out the trivial ones for which the corresponding Hamiltonians vanish. It turns out that the resulting algebra has a one-to-one correspondence with $\mathfrak{g}_\mathfrak{u}$\cite{CFP}.

\section{Boundary conditions and variational principle with null boundaries}
The variational principle for Einstein's general relativity with null boundaries has been investigated in \cite{PCMP,PCP,LMPS,JSSS}. Here we follow the computational strategy developed in \cite{LMPS} with minor improvements. The main novelty lies in the fact that for our later purpose we impose the boundary condition on null boundaries according to the boundary structure mentioned above, which is totally different from that taken before as in \cite{PCMP,PCP,LMPS,JSSS}\footnote{The boundary condition considered in \cite{CS}, which appeared online after the present paper, is also different from ours in the sense that it is weaker than ours.}.

To proceed, we would like to first fix the partial gauge such that the location of the null boundary $\mathcal{N}$ and its foliation by a selected cross-section $\mathcal{S}$ and $l^a$ are unchanged for all the metrics in the configuration space. Namely, we can have the following decomposition
\begin{equation}
g^{ab}=-l^an^b-l^bn^a+q^{ab}
\end{equation}
with $q^{ab}$ tangent to $\mathcal{S}$ for all the metrics under consideration and
\begin{equation}
\delta l_a=\delta a l_a, \quad \delta l^a=0,
\end{equation}
which further implies that
\begin{equation}
\delta n_a=\dbar b l_a, \quad \delta n^a=-\dbar b l^a-\delta a n^a+\dbar l^a
\end{equation}
with $\dbar l^a$ tangent to $\mathcal{S}$\footnote{Here $\dbar b$ and $\dbar l^a$ denote infinitesimal quantities, but not variations of something.}.
As a result, the variation of the metric can be expressed as
\begin{equation}\label{variationmetric}
\delta g^{ab}=2\dbar b l^al^b+\delta a (l^an^b+l^bn^a)-l^a\dbar l^b-l^b\dbar l^a+\delta q^{ab}.
\end{equation}

Now let us perform the variation of  the Einstein-Hilbert action
\begin{equation}
S=\int_\mathcal{M}\bm{L}=\int_\mathcal{M} (R-2\Lambda)\bm{\epsilon}^{(d)},
\end{equation}
which gives rise to
\begin{equation}
\delta S=\int_\mathcal{M}-(G^{ab}+\Lambda g^{ab})\delta g_{ab}\bm{\epsilon}^{(d)}-\int_\mathcal{N}v^al_a\overline{\bm{\epsilon}}^{(d-1)}
\end{equation}
where for our purpose all the other boundary terms are ignored on non-null boundaries and
\begin{eqnarray}
v^al_a&\;\widehat{=}\;&(g^{bc}l_a-l^b\delta^c{}_a)\delta C^a{}_{bc}\nonumber\\
&=&(l^bl^cn_a-l^bn^cl_a+q^{bc}l_a-l^bq^c{}_a)\delta C^a{}_{bc}
\end{eqnarray}
with $\delta C^a{}_{bc}=\frac{1}{2}g^{ad}(\nabla_b\delta g_{cd}+\nabla_c\delta g_{bd}-\nabla_d \delta g_{bc})$. On the other hand, by writing the surface gravity $\kappa=-n_bl^a\nabla_al^b=-n^bl^a\nabla_al_b$, we have
\begin{equation}\label{vkap}
\delta \kappa=-n_al^bl^c\delta C^a{}_{bc}, \quad \delta \kappa=l_al^bn^c\delta C^a{}_{bc}+l^a\nabla_a\delta a.
\end{equation}
In addition, by writing the expansion $\theta=q^{ab}\nabla_al_b=q^a{}_b\nabla_al^b$, we have
\begin{equation}\label{vthe}
\delta \theta=\nabla_al_b\delta q^{ab}+\theta \delta a-l_aq^{bc}\delta C^a{}_{bc},\quad \delta \theta=q^b{}_al^c\delta C^a{}_{bc},
\end{equation}
where we have used $q^a{}_b=\delta^a{}_b+l^an_b+n^al_b$ to perform the variation  for the second equation. With the help of Eq. (\ref{vkap}) and Eq. (\ref{vthe}), the boundary term can be expressed as
\begin{eqnarray}
&&-\int_\mathcal{N}v^al_a\overline{\bm{\epsilon}}^{(d-1)}\nonumber\\
&&=\int_\mathcal{N}(2\delta\theta -\nabla_al_b\delta q^{ab}+2\delta \kappa-\theta \delta a-l^a\nabla_a\delta a)\overline{\bm{\epsilon}}^{(d-1)}\nonumber\\
&&=\int_\mathcal{N}\left(2\delta\theta -\nabla_al_b\delta q^{ab}+2\delta \kappa-D_a(l^a\delta a)\right)\overline{\bm{\epsilon}}^{(d-1)}\nonumber\\
&&=\int_\mathcal{N}\left[\delta (2\theta\overline{\bm{\epsilon}}^{(d-1)})-\left((\nabla_al_b-\theta q_{ab})\delta q^{ab}-2\delta \kappa\right)\overline{\bm{\epsilon}}^{(d-1)}\right]\nonumber\\
&&-\int_{\partial\mathcal{N}}\delta a\overline{\bm{\epsilon}}^{(d-2)}\nonumber\\
&&=-\int_\mathcal{N}\left((\nabla_al_b-\theta q_{ab})\delta q^{ab}-2\delta \kappa\right)\overline{\bm{\epsilon}}^{(d-1)}\nonumber\\
&&+\int_{\partial\mathcal{N}}(2\delta\overline{\bm{\epsilon}}^{(d-2)}-\delta a\overline{\bm{\epsilon}}^{(d-2)}).
\end{eqnarray}
With the usual boundary condition $\delta q^{ab}\;\widehat{=}\;0$, the variational principle is well defined at the null boundaries as the timelike boundaries if the Einstein-Hilbert action is supplemented with a boundary term
\begin{equation}
-2\int_\mathcal{N}\kappa \overline{\bm{\epsilon}}^{(d-1)}+\int_{\partial\mathcal{N}}a\overline{\bm{\epsilon}}^{(d-2)},
\end{equation}
which is non-invariant under the reparametrization of the null generators. One can restore such an invariance by adding an extra counter term if one wants. Moreover, such a viable counter term is non-unique\cite{LMPS}.
However, the boundary condition we impose is different from this usual one. Instead, we work with the configuration space $\mathscr{F}_\mathfrak{p}$, whose induced boundary structure on $\mathcal{N}$ is $\mathfrak{p}=[l^a,\kappa,l_a]$. By the equivalence relation (\ref{triple}), $\delta l^a=0$ implies $\alpha=0$. This amounts to saying that to preserve the boundary structure, the boundary condition for the variation of the metric is required to satisfy
\begin{equation}
\delta\kappa\;\widehat{=}\;0, \quad \delta a\;\widehat{=}\;0.
\end{equation}
With this boundary condition, the resulting boundary term reads
\begin{eqnarray}\label{novel}
&&-\int_\mathcal{N}v^al_a\overline{\bm{\epsilon}}^{(d-1)}\nonumber\\
&&=\int_\mathcal{N}\left(\delta (2\theta\overline{\bm{\epsilon}}^{(d-1)})-(\nabla_al_b-\theta q_{ab})\delta q^{ab}\overline{\bm{\epsilon}}^{(d-1)}\right),
\end{eqnarray}
which exhibits a distinct feature from the previous case, because the second term cannot be written generically as a variation of something\footnote{The second term vanishes automatically for $d=2$ and $d=3$ because the hatted space is zero dimensional for $d=2$ and there is no shear for $d=3$. But nevertheless, we are interested in $d\geq 4$, where it does not vanish generically because $\delta q^{ab}\;\widehat{=}\;0$ is not our boundary condition any more.}. So the variational principle does not work in $\mathscr{F}_\mathfrak{p}$. However, by supplementing the Einstein-Hilbert action with the minus of the first term as
\begin{equation}\label{surface}
-2\int_\mathcal{N}\theta \overline{\bm{\epsilon}}^{(d-1)},
\end{equation}
the variational principle works in the submanifold of $\mathscr{F}_\mathfrak{p}$, denoted as $\mathscr{F}_\mathfrak{f}$, where free horizons serve as null boundaries.

\section{Harlow-Wu's algorithm and Wald-Zoupas' prescription with null boundaries\label{competition}}

\begin{figure}
\begin{center}
  \includegraphics[scale=0.52]{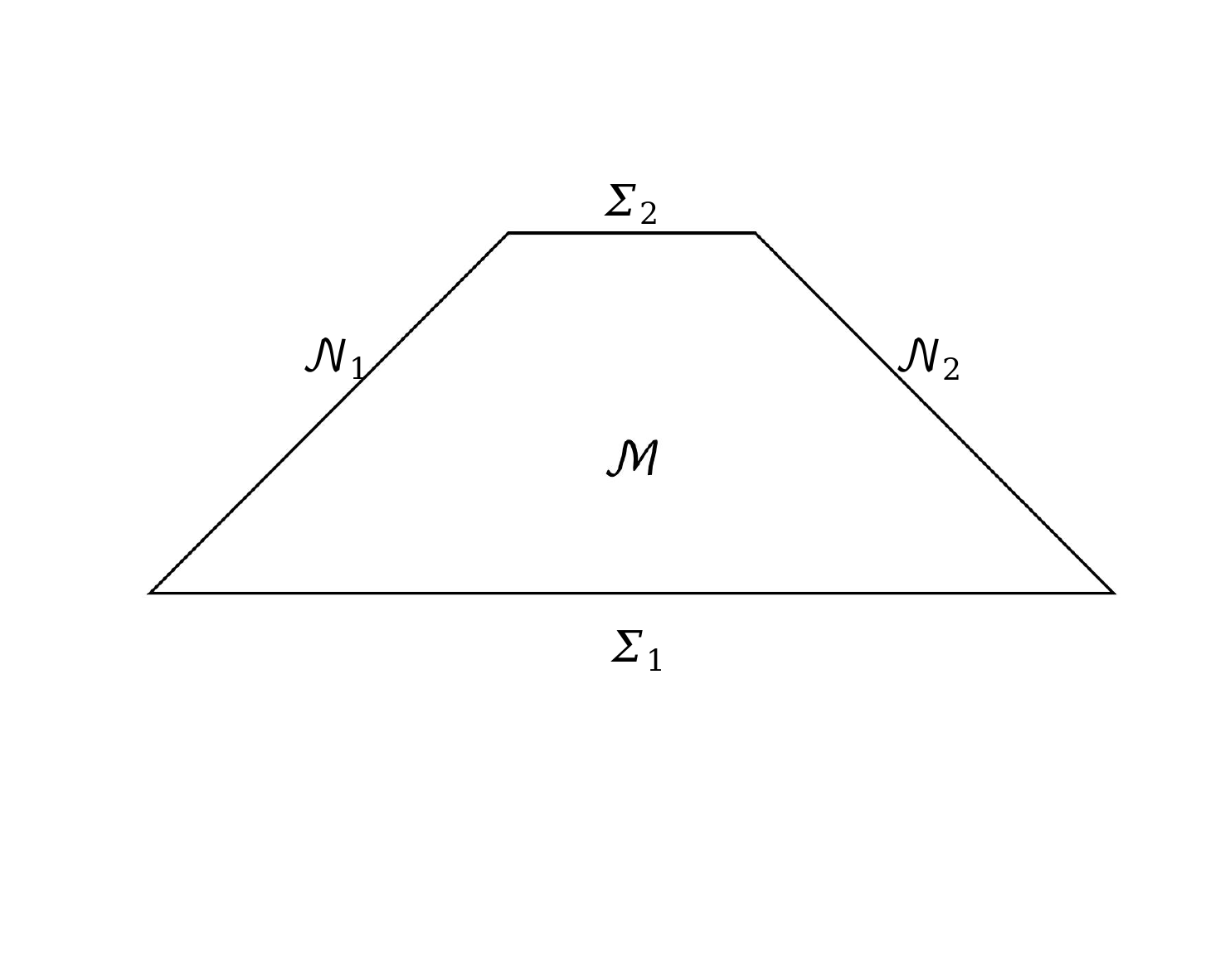}
  \end{center}
  \caption{The null hypersurface $\mathcal{N}_1$ is the inner boundary of the theory such as the black hole event horizon, and $\mathcal{N}_2$ is the outer boundary, whose limit can be thought of as the null infinity in an asymptotically flat spacetime.}\label{manifold}
\end{figure}

Let us first review  Harlow-Wu's algorithm by adapting it to a manifold $M$ with null boundaries for general covariant Lagrangian theories. For our purpose, the corresponding boundary is specified in Figure.\ref{manifold} as $\partial M=\mathcal{N}_1\cup\Sigma_1\cup\Sigma_2\cup\mathcal{N}_2$ with $\mathcal{N}_i$ null and $\Sigma_i$ spacelike,  which gives rise to the corner $\mathcal{S}_{ij}=\mathcal{N}_i\cap\Sigma_j$. To make the variational principle well defined in this circumstance, the corresponding action is generically required to be supplemented with the boundary condition dependent surface term as
\begin{equation}
S=\int_M \bm{L}+\int_{\partial M}\bm{B},
\end{equation}
where $\bm{B}$ is required to be covariant only on the boundary.
Although both $\Sigma_i$ and $\mathcal{N}_i$ are the portions of the boundary, they play different roles in the theory. In general, the data on $\Sigma_i$ correspond to the state of the theory while the data on $\mathcal{N}=\mathcal{N}_1\cup\mathcal{N}_2$ prescribe the boundary condition of the theory. In this sense, under the boundary condition we impose on $\mathcal{N}$, the variation of the on-shell action is not required to vanish, but instead is allowed to be of the following form\footnote{This form of the variation of the on-shell action amounts to saying that the variational principle works because it involves no boundary term on $\mathcal{N}_i$.}
\begin{equation}\label{onshell}
\delta S=\int_{\Sigma_2}\bm{\Psi}-\int_{\Sigma_1}\bm{\Psi},
\end{equation}
where the induced volume on $\Sigma$ is specified as $\bm{\epsilon}^{(d)}=\bm{t}\wedge\bm{\epsilon}^{(d-1)}$ with $\bm{t}$ the future directed normal vector to $\Sigma$.
This requires that
\begin{equation}\label{toostrong}
\bm{\Theta}+\delta\bm{B}\;\widehat{=}\;\text{d}\bm{C}
\end{equation}
where $\bm{\Theta}$ is  determined by
\begin{equation}
\delta \bm{L}=\bm{E}\delta \phi+\text{d}\bm{\Theta},
\end{equation}
and $\bm{C}$ is required to be covariant only on the boundary.
Whence it is not hard to show that
\begin{equation}
\bm{\Psi}|_{\Sigma_i}=\bm{\Theta}+\delta\bm{B}-\text{d}\bm{C},
\end{equation}
where we have used the fact that the orientation induced at the corner $S_{ij}$ $\partial\mathcal{N}_i$ by $\mathcal{N}_i$ is opposite to that at $\partial\Sigma_j$ induced by $\Sigma_j$ via the contraction of the outward pointing normal with the first index of the volume of $\Sigma_j$. Hereafter we shall view $\delta$ as the exterior derivative in the configuration space $\mathscr{F}$ under consideration and also denote the exterior derivative by $\delta$ on its submanifold such as the corresponding covariant phase space $\bar{\mathscr{F}}$, which is defined as the collection of the on-shell configurations in $\mathscr{F}$. With this in mind, the pre-symplectic current is then defined as
\begin{equation}
\bm{\omega}=\delta\bm{\Psi}=\delta(\bm{\Theta}-\text{d}\bm{C}),
\end{equation}
whereby the pre-symplectic form defined as
\begin{equation}
\Omega=\int_{\Sigma_i}\bm{\omega}
\end{equation}
is obviously $\Sigma_i$ independent on the covariant phase space $\bar{\mathscr{F}}$ because of Eq. (\ref{onshell}). Then associated with a vector field in the configuration space induced by the infinitesimal asymptotic symmetry generator $\xi$ as
\begin{equation}
X_\xi=\int d^dx\sqrt{-g}\mathscr{L}_\xi\phi(x)\frac{\delta}{\delta\phi(x)},
\end{equation}
one can introduce a Hamiltonian on $\mathscr{F}$ as
\begin{equation}\label{old}
\delta H_\xi(\Sigma_i)|_{\bar{\mathscr{F}}}=- X_\xi\cdot\Omega,
\end{equation}
which is well defined because one can show that
\begin{equation}\label{new}
\delta H_\xi(\Sigma_i)|_{\bar{\mathscr{F}}}=\delta\left(\int_{\Sigma_i} \bm{J}_\xi+\int_{\partial\Sigma_i}(\xi\cdot\bm{B}-X_\xi\cdot\bm{C})\right),
\end{equation}
where $\bm{J}_\xi$ is the Noether current, defined as
\begin{equation}
\bm{J}_\xi=X_\xi\cdot\bm{\Theta}-\xi\cdot\bm{L}.
\end{equation}
When restricted onto $\bar{\mathscr{F}}$, we have $\bm{J}_\xi=\text{d}\bm{Q}_\xi$ with $\bm{Q}_\xi$ the Noether charge. Thus one can integrate out the Hamiltonian  as
\begin{equation}\label{conserved}
H_\xi(\Sigma_i)|_{\bar{\mathscr{F}}}=\int_{\partial\Sigma_i}(\bm{Q}_\xi+\xi\cdot\bm{B}-X_\xi\cdot\bm{C})+\text{const},
\end{equation}
which turns out to be independent of the choice of $\Sigma_i$ because $\xi$ is tangent to $\mathcal{N}$.
\begin{figure}
\begin{center}
  \includegraphics[scale=0.42]{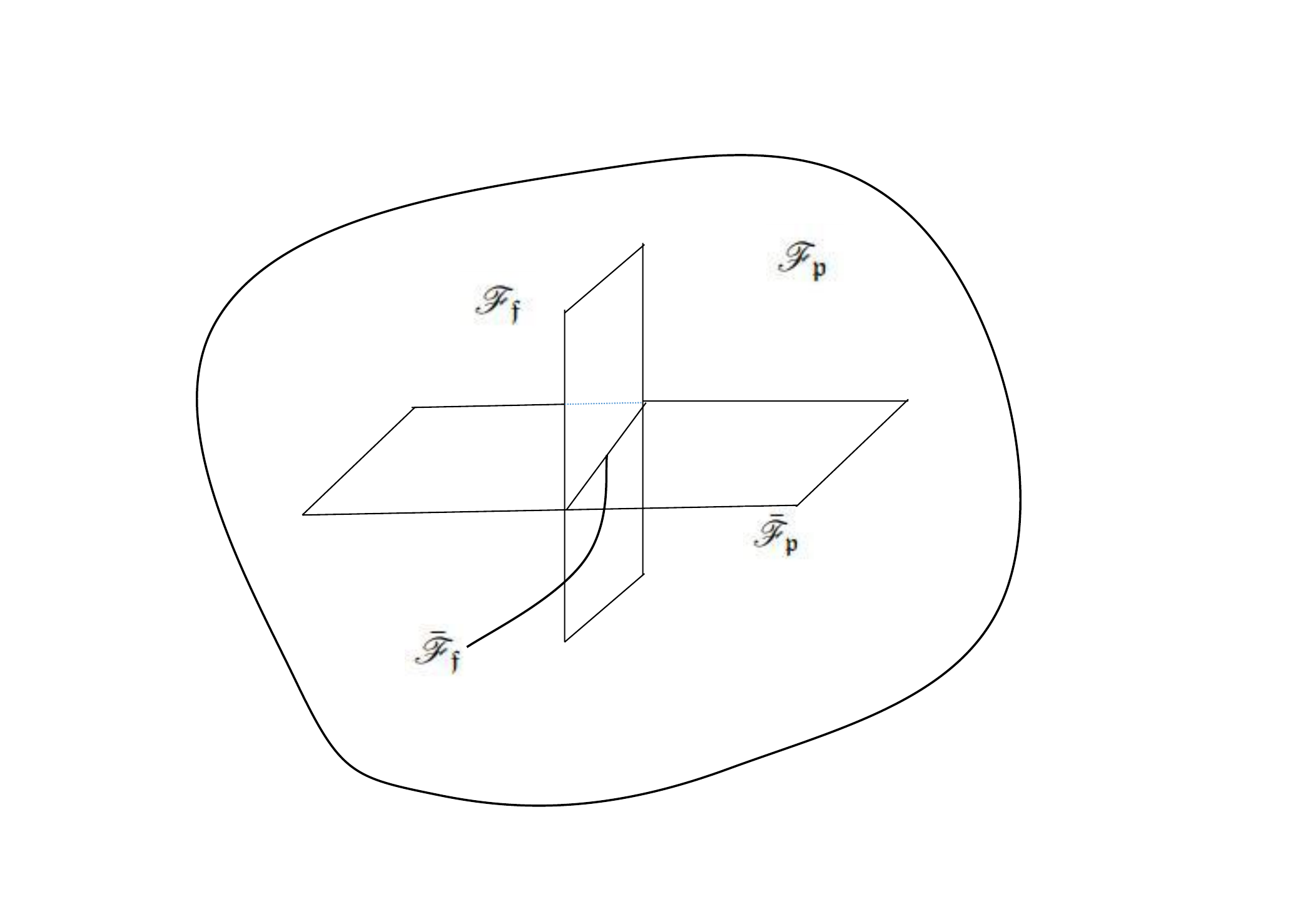}
  \end{center}
  \caption{$\mathscr{F}_\mathfrak{p}$ is specified as the configuration space associated with the null boundary structure, where $\bar{\mathscr{F}}_\mathfrak{p}$ represents the covariant phase space, while $\mathscr{F}_\mathfrak{f}$ denotes its submanifold with null boundaries given by free horizons. The intersection of $\bar{\mathscr{F}}_\mathfrak{p}$ and $\mathscr{F}_\mathfrak{f}$ gives rise to the submanifold as the stand point for the definition of the Hamiltonian.}\label{configurationspace}
\end{figure}

However, as alluded in Eq. (\ref{novel}), the requirement prescribed by Eq. (\ref{onshell}) together with Eq. (\ref{toostrong}) is so restrictive that it freezes the boundary dynamics in such a way that one fails to address such interesting phenomena as the gravitational waves passing through the boundary. In order to incorporate such a dynamics, we would like to relax the requirement as follows\footnote{Such a relaxation also applies to the case in which the boundary $\mathcal{N}$ is timelike.  In particular, for asymptotically locally Anti-de Sitter boundaries, such an issue has been systematically explored in \cite{PS}, where the conformal anomaly term in odd dimension is analogous to the $\bm{F}$ term here.}
\begin{equation}
\bm{\Theta}+\delta\bm{B}\;\widehat{=}\;\text{d}\bm{C}+\bm{F},\quad \delta S=\int_{\Sigma_2}\bm{\Psi}-\int_{\Sigma_1}\bm{\Psi}+\int_\mathcal{N}\bm{F}
\end{equation}
in the whole covariant phase space. When restricted onto its submanifold on which $\bm{F}$ vanishes, one has a well defined variational principle and can construct the conserved Hamiltonian by following the procedure reviewed above. On the other hand, with the presence of the additional $\bm{F}$ term off the aforementioned submanifold, the variational principle is ill defined, but one has no obstruction to take Eq. (\ref{conserved}) as the definition of the Hamiltonian, which is nevertheless not conserved in general, as it should be the case. Instead, the change of the Hamiltonian can be evaluated in terms of the flux across the null boundary $\mathcal{N}$ as
\begin{eqnarray}\label{change}
&&H_\xi(\Sigma_2)-H_\xi(\Sigma_1)\nonumber\\
&&=\int_\mathcal{N}\left(\bm{J}_\xi+\text{d}(\xi\cdot\bm{B}-X_\xi\cdot\bm{C})\right)\nonumber\\
&&=\int_\mathcal{N}\left(X_\xi\cdot\bm{\Theta}-\xi\cdot \bm{L}-\xi\cdot \text{d}\bm{B}+\mathscr{L}_\xi\bm{B}-X_\xi\cdot \text{d}\bm{C}\right)\nonumber\\
&&=\int_\mathcal{N}\left(X_\xi\cdot(\bm{\Theta}-\text{d}\bm{C})+\mathscr{L}_{X_\xi}\bm{B}\right)\nonumber\\
&&=\int_\mathcal{N}X_\xi\cdot(\bm{\Theta}+\delta\bm{B}-\text{d}\bm{C})=\int_\mathcal{N}X_\xi\cdot \bm{F},
\end{eqnarray}
where we have used the fact that $\xi^a$ is tangent to $\mathcal{N}$ in the second step, the covariance of $\bm{B}$ in the third step, $X_\xi\cdot \bm{B}=0$ in the fourth step. In addition, it is noteworthy that Eq. (\ref{old}) and Eq. (\ref{new}) are generically not equal to each other in the presence of the $\bm{F}$ term. The corresponding difference can also be calculated in a general way as
\begin{eqnarray}\label{different}
&&X_\xi\cdot\Omega+\delta H_\xi(\Sigma_i)\nonumber\\
&&=\int_{\Sigma_i} \left(X_\xi\cdot\delta (\bm{\Theta}-\text{d}\bm{C})+\delta (X_\xi\cdot\bm{\Theta}-\xi\cdot \bm{L})\right)\nonumber\\
&&+\int_{\partial\Sigma_i}\delta (\xi\cdot\bm{B}-X_\xi\cdot \bm{C})\nonumber\\
&&=\int_{\Sigma_i}(\mathscr{L}_{X_\xi}\bm{\Theta}-\xi\cdot\delta \bm{L})+\int_{\partial\Sigma_i}(\xi\cdot\delta\bm{B}-\mathscr{L}_{X_\xi}\bm{C})\nonumber\\
&&=\int_{\Sigma_i}(\mathscr{L}_\xi\bm{\Theta}-\xi\cdot \text{d}\bm{\Theta})+\int_{\partial\Sigma_i}(\xi\cdot \delta\bm{B}-\mathscr{L}_\xi\bm{C})\nonumber\\
&&=\int_{\partial\Sigma_i}\xi\cdot(\bm{\Theta}+\delta\bm{B}-\text{d}\bm{C})=\int_{\partial\Sigma_i}\xi\cdot\bm{F}.
\end{eqnarray}

By holding both $\bm{E}$ and $\bm{F}$ fixed, all the remaining ambiguities in the course of the definition of the above Hamiltonian can be spelt out as follows
\begin{eqnarray}\label{unique}
&&\bm{L}\rightarrow \bm{L}+\text{d}\bm{T},\nonumber\\
&&\bm{\Theta}\rightarrow \bm{\Theta}+\text{d}\bm{Y}+\delta \bm{T},\nonumber\\
&&\bm{B}\rightarrow \bm{B}+\text{d}\bm{D}-\bm{T},\nonumber\\
&&\bm{C}\rightarrow \bm{C}+\text{d}\bm{K}+\delta \bm{D}+\bm{Y},\nonumber\\
&&\bm{J}_\xi\rightarrow\bm{J}_\xi+\text{d}(X_\xi\cdot\bm{Y}+\xi\cdot\bm{T}),\nonumber\\
&&\bm{Q}_\xi\rightarrow \bm{Q}_\xi+\text{d}\bm{G}+X_\xi\cdot\bm{Y}+\xi\cdot\bm{T}, \nonumber\\
&&H_\xi\rightarrow H_\xi+\text{const}.
\end{eqnarray}
This tells us that the only ambiguity for the final Hamiltonian comes from the aforementioned integral constant.
In this sense, the resulting Hamiltonian is determined uniquely up to the integral constant by $\bm{E}$ and $\bm{F}$. This is reasonable because the former gives rise to the covariant phase space and the latter gives rise to its submanifold on which $\bm{F}$ vanishes as the stand point for the definition of the Hamiltonian. The canonical choice of $\bm{F}$ is such that $\bm{F}$ involves no variation of the derivatives of the metric.

With the above preparation, now let us focus on Einstein's general relativity. According to Eq. (\ref{novel}), the canonical choice of $\bm{F}$ is given by
\begin{equation}
\bm{F}=-(\nabla_al_b-\theta q_{ab})\delta q^{ab}\overline{\bm{\epsilon}}^{(d-1)},
\end{equation}
and
\begin{equation}
\bm{B}=-2\theta\overline{\bm{\epsilon}}^{(d-1)}, \quad \bm{C}=0,
\end{equation}
where our canonical choice of $\bm{F}$ is reasonable as it vanishes when evaluated in $\mathscr{F}_\mathfrak{f}$. As a result, the Hamiltonian constructed by Eq. (\ref{conserved}) is exactly the same as that obtained in \cite{CFP} by Wald-Zoupas' prescription, which is a conserved quantity on $\bar{\mathscr{F}}_\mathfrak{f}$, consisting of on-shell configurations with free horizons as null boundaries, depicted in Figure.\ref{configurationspace}. However, by Eq. (\ref{change}), we have
\begin{eqnarray}
&&H_\xi(\Sigma_2)-H_\xi(\Sigma_1)\nonumber\\
&&=-\int_\mathcal{N}X_\xi\cdot(\nabla_al_b-\theta q_{ab})\delta q^{ab}\overline{\bm{\epsilon}}^{(d-1)}\nonumber\\
&&=-\int_\mathcal{N}X_\xi\cdot(\nabla_al_b-\theta q_{ab})\delta g^{ab}\overline{\bm{\epsilon}}^{(d-1)}\nonumber\\
&&=2\int_\mathcal{N}(\nabla_al_b-\theta q_{ab})\nabla^a\xi^b\overline{\bm{\epsilon}}^{(d-1)}
\end{eqnarray}
for a generic null boundary, where Eq. (\ref{variationmetric}) is used  in the second step. By Eq. (\ref{different}), we also have
\begin{equation}
X_\xi\cdot\Omega+\delta H_\xi=-\int_{\partial\Sigma}(\nabla_al_b-\theta q_{ab})\delta g^{ab}\xi\cdot\overline{\bm{\epsilon}}^{(d-1)}.
\end{equation}


\section{Conclusion and discussions}
Guided by the variational principle, Harlow-Wu's algorithm provides us with a routine way to define the Hamiltonians associated with the infinitesimal asymptotic symmetries by incorporating all the boundary terms in a meticulous way. In this sense, Harlow-Wu's algorithm improves on Wald-Zoupas' prescription, which involves well educated guess somehow. By imposing the boundary condition associated with the boundary structure of the null boundaries rather than the usual one, we find for Einstein's gravity that the variational principle works only in its submanifold rather than the whole covariant phase space, which implies that the key requirement in Harlow-Wu's algorithm for timelike boundaries is too restrictive for null boundaries. As such, we relax such a requirement to incorporate more generic situations into Harlow-Wu's algorithm. With such a relaxation, we successfully reproduce the Hamiltonian obtained previously in \cite{CFP} by Wald-Zoupas' prescription for Einstein's gravity.

There are a variety of issues worthy of further investigation.  For instance, although the null infinity in an asymptotically flat spacetime can be viewed as a limit of $\mathcal{N}_2$ for example, it is better to work within the conformal completion framework, where some new subtleties may arise in reproducing Wald-Zoupas' prescription by Harlow-Wu's algorithm. In addition, here we focus only on Einstein's gravity. As argued in \cite{WZ}, the Hamiltonians associated with the null infinity are supposed to be the same in higher derivative gravity theories as in Einstein's general relativity. However, it is expected that the Hamiltonians will acquire some additional corrections in higher derivative gravity theories when the null boundary is located at a finite position. So it is intriguing to obtain the corresponding Hamiltonians by Harlow-Wu's algorithm. We expect to report both of these issues somewhere else in the near future.

\section*{Acknowledgement}
H.Z. is grateful to Bob Wald for his helpful correspondence related to this project. We also like to thank Jie Jiang for his useful discussions.
This work is partially supported by the National Science Foundation of China under the Grant No.11675015, 11875095, and 12075026. In addition, H.Z. is supported in part by FWO Vlaanderen through the project G006918N, and by the Vrije Universiteit Brussel through
the Strategic Research Program ``High-Energy Physics''. He is also an individual FWO
Fellow supported by 12G3518N.

\end{document}